# Use of Emergency Departments by Frail Elderly Patients: Temporal Patterns and Case Complexity

Jens RAUCH[a,1], Mathias DENTER[b] and Ursula HÜBNER[a]

[a] *Health Informatics Research Group, Osnabrück University AS, Germany*
[b] *Klinikum Osnabrück, Germany*

**Abstract.** Emergency department (ED) care for frail elderly patients is associated with an increased use of resources due to their complex medical needs and frequently difficult psycho-social situation. To better target their needs with specially trained staff, it is vital to determine the times during which these particular patients present to the ED. Recent research was inconclusive regarding this question and the applied methods were limited to coarse time windows. Moreover, there is little research on time variation of frail ED patients' case complexity. This study examines differences in arrival rates for frail vs. non-frail patients in detail and compares case complexity in frail patients within vs. outside of regular GP working hours. Arrival times and case variables (admission rate, ED length of stay [LOS], triage level and comorbidities) were extracted from the EHR of an ED in an urban German teaching hospital. We employed Poisson time series regression to determine patterns in hourly arrival rates over the week. Frail elderly patients presented more likely to the ED during already high frequented hours, especially at midday and in the afternoon. Case complexity for frail patients was significantly higher compared to non-frail patients, but varied marginally in time only with respect to triage level and ED LOS. The results suggest that frailty-attuned emergency care should be available in EDs during the busiest hours. Based on EHR data, hospitals thus can tailor their staff needs.

## 1. Introduction

One of the most pertinent issues for emergency departments (ED) is the growing number of frail elderly patients, who seek emergency treatment [1]. Frailty, which progresses with increasing age, refers to a functional decline in several organ systems [2]. Not only do these patients typically present with more comorbidities, also they often show unusual and non-specific symptoms or delirium, which complicates diagnoses and treatment [3]. Leading to a higher case complexity, frail patients therefore tie up considerable resources, especially as to date EDs are commonly not set up for specialised geriatric medicine. There is no doubt that targeted intervention by a multi-disciplinary team of trained professionals will result in a better clinical outcome for these patients [4]. Since the provision of such specialised teams involves increased expenditure, it needs to be determined when frail patients typically present to the ED over the course of the day and week. Recent studies concerning patterns in time were inconclusive regarding the question, whether frail patients rather present within or outside of regular GP working hours [5,6]. A detailed analysis of hourly presentations of frail patients over the course of a week has not been published, so far. Moreover, analyses of case complexity have only been conducted for the entire study period or days of the week, disregarding potential dependencies in time of day.

[1] Corresponding Author: Jens Rauch, Osnabrück University of AS, Health Informatics Research Group, PO Box 1940, 49009 Osnabrück, Germany; E-Mail: j.rauch@hs-osnabrueck.de.

The aim of the present study is thus to identify detailed temporal patterns in ED use and case complexity of frail patients. Specifically, we pose the following research questions: Do arrival times of frail patients deviate from non-frail ED cases and time of day or day of week are there particular differences? Do these patients rather present outside regular GP working hours? How does medical complexity of frail patients vary over the day and how does its time course compare to other patients' complexity?

## 2. Methods

We conducted a retrospective study and used historical ED data extracted from the Electronic Health Record of Klinikum Osnabrück, an academic teaching hospital with 660 beds serving the town and region of Osnabrück, Lower Saxony, Germany. The ED has about 40,000 cases per year and is operated 24 hours a day on 365 days a year. Data was being captured in a custom ED module of the clinical information system Cerner Medico. Since exported data was anonymised no ethical approval was required [7]. Data covered the period from January 1, 2017 until July 31, 2018. We employed a data cleansing process that removed duplicate data sets, visits with non-positive or excessive LOS (> 10 h). Further we included only patients that were at least 18 years old and had at least one diagnosis assigned. Frail patients were identified as patients who were 75 years and older and had at least 5 points on the Hospital Frailty Risk Score (HFRS) [2]. The HFRS is solely based on ICD-10 codings and is designed to systematically identify groups of patients from routine EHR data for whom a specific frailty-attuned approach should be considered. Only admission and discharge ICD codes were used. To determine the degree of case complexity [8,9], we examined the following variables: assigned triage code by the Manchester Triage System (MTS) [10], whether the patient was admitted as an in-patient and the weighted score of the Charlson comorbidity index [11]. In addition, we analysed the length-of-stay (LOS) within the ED. For temporal analysis, we aggregated data into one-hour bins spanning the entire study period. Subsequently, we calculated total hourly presentations and centrality measures for all variables of interest for both frail and all other patients. Since arrivals to the ED typically show a distinct weekday pattern [12], analysis was conducted depending on day of week and hour of day, yielding 24∗7 = 168 hour slots. The normalised hourly arrival rate $\rho$ per slot was calculated by averaging arrival counts of a slot and dividing by the total number of patients in the respective group (frail vs. non-frail) over the entire period of study. Each hourly arrival distribution was assumed to be homogeneously Poisson distributed. To compare arrival counts of patient-groups, we therefore fitted a Poisson time series regression model [13] with separate approximations for day-of-week and hour-of-day cycles. The regression model was specified as in equation (1).

$$\ln E(n_t) = \ln N + a_0 + \sum_{k=1}^{K_w}\left[a_k \cos\left(\frac{2\pi k d(t)}{7}\right) + b_k \sin\left(\frac{2\pi k d(t)}{7}\right)\right]$$
$$+ \sum_{k=1}^{K_d}\left[\alpha_k \cos\left(\frac{2\pi k h(t)}{24}\right) + \beta_k \sin\left(\frac{2\pi k h(t)}{24}\right)\right]$$
$$+ \mathbf{1}_{isfrail}\left(\sum_{k=1}^{K_w}\left[x_k \cos\left(\frac{2\pi k d(t)}{7}\right) + y_k \sin\left(\frac{2\pi k d(t)}{7}\right)\right]\right.$$
$$\left.+ \sum_{k=1}^{K_d}\left[\xi_k \cos\left(\frac{2\pi k h(t)}{24}\right) + \eta_k \sin\left(\frac{2\pi k h(t)}{24}\right)\right]\right) \quad (1)$$

Here $n_t$ is the number of ED arrivals, $d(t)$ the day of week and $h(t)$ the hour of day at hour $t$. The number of Fourier terms $K_w$ and $K_d$ to be included were iteratively determined by the Akaike Information Criterion (AIC). We added $ln(N_p)$ ($N$ being the total number of patients in the cohort for each group) as an offset term to account for different overall arrival rates in the groups. The model was initially fitted for non-frail patients only. In the next step, we analysed whether there was a difference in arrival patterns for frail patients. To this end, we included frail patients and tested, whether the model could be improved by two additional Fourier series terms (with coefficients $x_k, y_k$ for day-of-week and $\xi_k, \eta_k$ for hour-of-day) of the same form as above. These additional series were set to zero for non-frail patients. We tested for a better model fit by comparing AIC and ANOVA of the regression model with versus without additional frail Fourier series. For admission rate, triage level and comorbidity as measures of case complexity, we performed Mann-Whitney U tests to compare metrics within and outside of regular GP working hours (Mondays through Fridays, 7 a.m. to 5 p.m.), as well as to compare frail and other patients. LOS was compared by estimating Restricted Mean Survival Time (RMST), which is the preferred option when the proportional hazards assumption cannot be guaranteed [14].

**Table 1.** Number of cases during the period of study.

|  | Non-frail | Frail | Total |
|---|---|---|---|
| Within GP hours | 20,693 | 3,152 | 23,845 |
| Outside GP hours | 21,399 | 2,837 | 24,272 |
| Total | 42,092 | 6,025 | 48,117 |

## 3. Results

In total, 13,451 elderly patients (≥ 75 y) presented to the ED during the period of study, 44.8% ($n = 6,025$) of which qualified as frail according to the HFRS criterion (Table 1). The proportion of frail elderly patients to the overall ED sample ($n = 48,117$) amounted to 12%. Arrival rates for both groups were highest on Mondays and decreased over the week, except for Fridays, which showed the second-highest arrival rates (Fig. 1.A). Generally, arrivals rates were lower during the weekend. The shapes arrival rates took over a day were similar for both groups and all weekdays, showing a daily peak around midday and a second peak in the afternoon. However, these peaks were much more pronounced for frail patients, than for other patients, during the week. On Saturdays, normalized arrivals rates matched over the day, while on Sundays frail patients tended to present less. Comparison of expected and actual arrivals for both frail and non-frail patients showed no severe violations of the assumption of a homogeneous Poisson distribution. A quasi-Poisson model gave an estimated dispersion parameter of 1.08 for non-frail patients and 1.01 for frail patients, which we deemed negligible. Iterative determination of the optimal number of Fourier terms to be included resulted in $K_w = 3$ and $K_d = 7$. Results of the Poisson time series regression analysis are given in Table 2. Analysis revealed that the model, which included an additional Fourier series term for frail patient arrivals, was superior compared to the model, which did not account specifically for frail patients. This was evident from both a lower AIC (73,048, $df = 21$ to 73,018, $df = 42$) and based on ANOVA (Table 3). Figure 1.A presents model fits for frail and other patients. The vast majority (91.6%) of frail patients was admitted to the hospital after treatment in the ED. This differed to the ratio of non-frail patients (46.7%, $\chi^2 = 4,260$, $p < .001$). There were no differences in admission rates w.r.t. time of arrival for frail patients (92.1% vs. 91.1%; $\chi^2 = 1.6$, $p = .20$). Median triage for frail patients was three, thereby one level higher than for other patients ($p < .001$), while there was no median difference when comparing frail patients during the day with those during out of GP hours.

**Table 2.** Results from Poisson time series regression. All time-of-day coefficients in the extended model except $\xi_1$ and $\eta_1$ were non-significant and therefore omitted. *: $p < .05$, **: $p < .001$.

| Initial model | | | | Geriatric extension | | | |
|---|---|---|---|---|---|---|---|
| Coeff. | Est. | Coeff. | Est. | Coeff. | Est. | Coeff. | Est. |
| $a_0$ | −9.74** | | | $a_0$ | −9.74** | | |
| Sine terms | | Cosine terms | | Sine terms | | Cosine terms | |
| $a_1$ | .025** | $b_1$ | −.013* | $a_1$ | .021* | $b_1$ | −.007 |
| $a_2$ | −.024** | $b_2$ | −.026** | $a_2$ | −.034** | $b_2$ | −.022* |
| $a_3$ | −.021** | $b_3$ | −.008 | $a_3$ | −.029** | $b_3$ | −.006 |
| $\alpha_1$ | −.687** | $\beta_1$ | −.602** | $\alpha_1$ | −.678** | $\beta_1$ | −.592** |
| $\alpha_2$ | −.203** | $\beta_2$ | .269** | $\alpha_2$ | −.201** | $\beta_2$ | .266** |
| $\alpha_3$ | .129** | $\beta_3$ | −.079** | $\alpha_3$ | .122** | $\beta_3$ | −.077** |
| $\alpha_4$ | −.055** | $\beta_4$ | −.028** | $\alpha_4$ | −.058** | $\beta_4$ | −.025** |
| $\alpha_5$ | −.001 | $\beta_5$ | .044** | $\alpha_5$ | −.002 | $\beta_5$ | .047** |
| $\alpha_6$ | .024* | $\beta_6$ | −.024* | $\alpha_6$ | .027* | $\beta_6$ | −.019* |
| $\alpha_7$ | −.013 | $\beta_7$ | −.029** | $\alpha_7$ | −.013 | $\beta_7$ | −.027** |
| | | | | $x_1$ | .027 | $y_1$ | −.046** |
| | | | | $x_2$ | .077** | $y_2$ | −.029* |
| | | | | $x_3$ | .058* | $y_3$ | −.020 |
| | | | | $\xi_1$ | −.076* | $\eta_1$ | −.086** |

**Table 3.** ANOVA results of time series model comparison.

| | Resid. Df. | Resid. Dev | Deviance | p ($\chi$2) |
|---|---|---|---|---|
| Initial model | 27,483 | 28,748 | | |
| Extended model | 27,462 | 28,676 | 71.64 | < .001 |

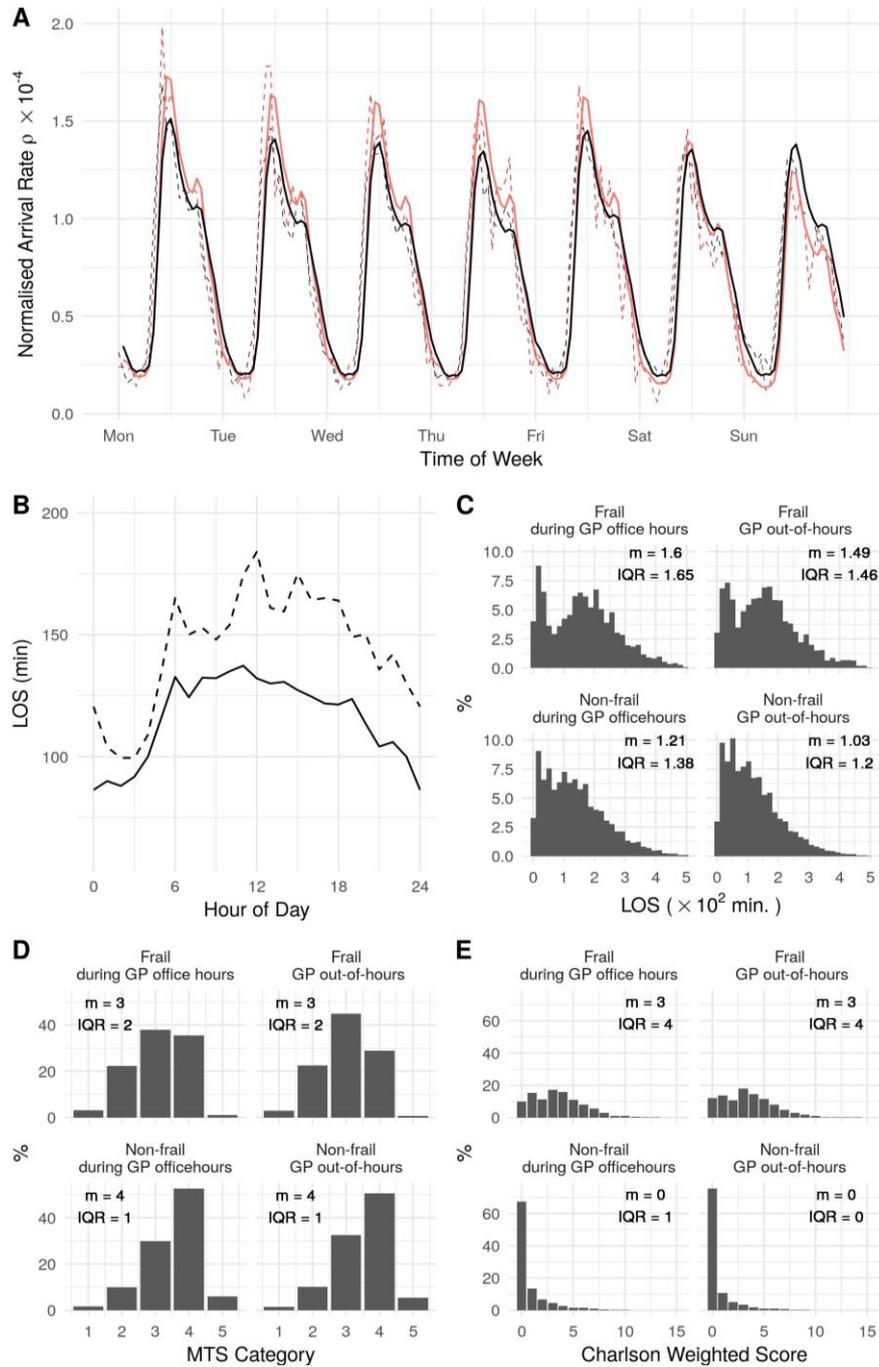

**Figure 1.** Comparison of frail vs. non-frail patients by time of day. (A) Modelled (solid lines) and observed (dashed lines) normalised mean arrival rates to the ED (black: non-frail patients). (B) Median length of stay in the ED dependent on time of arrival (dashed line: frail patients). (C-E) Relative frequencies and medians (*m*) and interquartile range (IQR) for LOS, triage (Manchester Triage System) and Charlson comorbidity score.

Charlson comorbidity score was higher in frail patients ($p < .001$, 95% lower bound estimate 3.0 score difference) but showed no dependence on time-of-day for frail patients ($p = 1$). Despite notable variations in overall median LOS over the day (Fig 1.B), RMST revealed a slightly longer LOS during the day ($p = .001$, 95% CI 4.0–14.3 LOS difference in min). Again, there was a conspicuous overall difference in LOS between frail and non-frail patients of about half an hour ($p < .001$, 95% CI 27.9–30.6 LOS difference in min).

## 4. Discussion

This study is the first to model and compare frail elderly patient arrival times with those of ordinary patients in emergency departments utilizing a Poisson time series regression. The resulting model confirms the two peaks in arrival time that have been reported consistently in various studies for the overall ED population [12,15]. Comparing frail and non-frail patients, we could identify notable differences between the two groups. The two arrival rate peaks on weekdays were more pronounced for frail patients than for all other patients. In the evening, at night and on weekends, relative presentation rates of frail patients in the ED were less than for typical non-frail patients. Previous studies in patterns of ED use by frail patients had made less detailed assertions regarding arrival rates. Some reported that frail patients present more likely during daytime hours [16,17]. However, the results are inconsistent specifically for nursing home residents: While one study [5] found fewer ED transfers outside of regular GP working hours, another one [6] reported a disproportionate large number of transfers at night and on weekends. Our study provides further evidence that frail patients tend to present within regular GP working hours. In addition, it provides new insight into the precise differences of hourly arrival rates between frail and non-frail patients, i.e. frail patients do much more likely present during midday or afternoon, while being even less likely to present on Sundays. Our time series model is able to describe patterns in ED arrivals that are common for all patients, such as weekday peaks on Mondays and Fridays, a two-peaked course over the day and lowest arrival rates during the weekends and night hours. Additionally, it particularly allows the arrivals of the resource-intensive frail patients to be modelled. There is a body of research devoted to day and week peaks in the occurrence of specific medical emergencies such as myocardial infarction and ischemic stroke [18]. These patterns are mostly attributed to physiological mechanisms which follow distinctive circadian patterns. Our results suggest that these well-known patterns might be even more manifest for frail patients. However, it is unclear what causes this pattern.

With respect to case complexity, we found significant higher levels for all examined variables (admission rate, ED LOS, MTS category, Charlson score) in frail patients, thereby confirming, what is known from the literature [19,20]. We did, however, only find a marginal relationship between case complexity and the time-of-day, with frail patients being assigned slightly higher triage levels outside and somewhat longer ED LOS during GP regular hours. Admission rate and comorbidities proved to be independent of time-of-day: frail patients were assigned slightly higher triage levels outside GP working hours and stayed longer in the ED during GP regular hours. While there are indeed considerable differences in the incidence of medical emergencies amongst the frail elderly over the day, the severity of which thus shows little variation. Pertaining to the discussion about potential inappropriate ED use by the elderly [6,21], it follows from our results that ED services are not misused for treatment of minor conditions neither during nor outside of regular GP working hours. This is in accordance with other findings that elderly patients mostly use EDs appropriately [17] and complements findings about stable degrees of case complexity over the week amongst elderly ED patients [22].

Our study is limited in that no data was present about an actual geriatric assessment of patients. While the FRHS provides a robust indicator for frailty, it is of course highly dependent on the quality a hospital's ICD documentation. Moreover, this is a single centre study and there could be differences in the arrival time curves in different EDs due to regional factors. This may also result in differences in the distribution and rate of frail patients. Also, close inspection of LOS for frail patients revealed a bimodal distribution (Fig. 1.C). This may indicate the presence of two sub-groups of frail patients: (1) patients with a clear medical indication and immediate transfer to surgery, and (2) patients with unclear symptoms and extensive ED diagnostics, as was reported as characteristic for many frail cases [19,6]. These patients consume more resources and should be examined further.

## 5. Conclusion

We found a significant difference in emergency department arrival rates for frail patients when compared to all other patients. Frail patients are more likely to arrive during the two arrival peaks over the day, that were commonly observed in EDs. They are less likely to arrive during GP out-of-office-hours and during the weekend. We could confirm considerable higher case complexity for frail patients in general, but were unable to find time dependent variations therein, apart from a somewhat less pronounced case acuity and longer ED LOS during the day. The findings suggest that trained intervention teams for frail patients in EDs are primarily needed during the day to cushion the already high demand for ED services.

## 6. Acknowledgements

We thank the Klinikum Osnabrück for the provision of emergency data and their collaboration. This work is funded by the state of Lower Saxony, project ROSE, the learning healthcare system (ZN 3103).